\documentclass[12pt]{article}

\usepackage{latexsym,bm,graphicx,color,xcolor,nicefrac,titletoc,enumerate,amsmath,amssymb,xfrac,xcolor,physics,cite,setspace,physics}

\usepackage{times}

\usepackage[hang,flushmargin,bottom]{footmisc} 

\usepackage[nottoc]{tocbibind} 

\usepackage{hyperref}
\hypersetup{linktocpage=true,colorlinks=true,linkcolor=blue,citecolor=blue,urlcolor=blue}

\usepackage[skip=3pt plus1pt, indent=20pt]{parskip}

\usepackage{geometry}
\usepackage{titlesec}

\titleformat{\section}{\large\bfseries}{\thesection}{0.5em}{}
\titleformat{\subsection}{\normalfont\bfseries}{\thesubsection}{0.5em}{}
\titleformat{\subsubsection}{\normalsize\itshape}{\thesubsubsection}{0.5em}{}
\titleformat*{\paragraph}{\normalsize\bfseries}

\numberwithin{equation}{section}

\def\a{\alpha}
\def\b{\beta}

\def\d{\delta}
\def\D{\Delta}
\def\e{\epsilon}

\def\f{\phi}
\def\vf{\varphi}

\def\l{\lambda}
\def\L{\Lambda}
\def\m{\mu}
\def\n{\nu}
\def\r{\rho}
\def\s{\sigma}

\def\th{\theta}

\def\x{\xi}
\def\z{\zeta}

\def\pd{\partial}

\def\pr{\prime}

\def\nn{\nonumber}
\def\qq{\quad\quad}

\newcommand{\rp}{r_+}

\newcommand{\be}{\begin{eqnarray}}
\newcommand{\ee}{\end{eqnarray}}

\newcommand{\cC}{\mathcal{C}}
\newcommand{\cD}{\mathcal{D}}
\newcommand{\cE}{\mathcal{E}}
\newcommand{\cF}{\mathcal{F}}
\newcommand{\cG}{\mathcal{G}}
\newcommand{\cH}{\mathcal{H}}

\newcommand{\cL}{\mathcal{L}}

\newcommand{\tT}{\widetilde{T}}

\newcommand{\bR}{\bar{R}}

\newcommand{\bnab}{\bar{\nabla}}

\newcommand{\dif}{\mathrm{d}}

\newcommand{\mail}[1]{\href{mailto:#1}{{\tt #1}}}

\begin{document}
	
\begin{titlepage}       
	\vspace{5pt} \hfill 
		
		\vspace{10pt}
		
	\begin{center}
		{\Large \bf Weyl double copy in Lifshitz spacetimes}
	\end{center}
		
		\begin{center}
			\vspace{10pt}
			
			{G{\"o}khan Alka\c{c}${}^{a}\,$, Mehmet Kemal G\"{u}m\"{u}\c{s}${}^{a}\,$ and Mehmet Ali Olpak}${}^{b}$
			\\[4mm]
			
			{\small 
				{\it ${}^a${Department of Aerospace Engineering, Faculty of Engineering,\\ At{\i}l{\i}m University, 06836 Ankara, T\"{u}rkiye\\[2mm]}}
				
				
				{\it ${}^b$Department of Electrical and Electronics Engineering, Faculty of Engineering,\\ University of Turkish Aeronautical Association, 06790, Ankara, Turkey}\\[2mm]
				
				{e-mail:} {\mail{alkac@mail.com}, \mail{kemal.gumus@metu.edu.tr}, \mail{maolpak@thk.edu.tr}}
			}
			\vspace{2mm}
		\end{center}

		\centerline{{\bf Abstract}}
		\vspace*{1mm}
		\noindent Lifshitz black hole solutions pose particular challenges for reconciling the two main formulations of the classical double copy: the Kerr-Schild double copy and the Weyl double copy. Recent work has suggested that consistency between the two can be restored, in certain cases, only by adopting a regularization prescription in the Weyl double copy. In this paper, we test this prescription on three examples from the literature, each with a distinct novel feature, and show that the prescription remains valid in all cases.
		\par\noindent\rule{\textwidth}{0.5pt}
		\tableofcontents
		\par\noindent\rule{\textwidth}{0.5pt}
\end{titlepage}
\section{Introduction}
String theory, our most prominent candidate for a quantum theory of gravity, has gifted us two astonishing relations between gravity and gauge theories. The AdS/CFT correspondence is a duality between gravity theories defined on AdS$_{D}$ spacetime and $(D-1)$-dimensional conformal field theories (CFTs) in the large $N$ limit  \cite{Maldacena:1997re,Witten:1998qj,Gubser:1998bc}. Thanks to the weak/strong nature of the duality, it is one of the most powerful tools for studying the nature of strongly coupled gauge theories. A second relation between gravity and gauge theories originates from a remarkable observation of Kawai, Lewellen and Tye regarding string amplitudes. At the tree-level, any closed string amplitude can be written as a sum of the products of certain open string amplitudes \cite{Kawai:1985xq}. As shown by Bern, Carrasco and Johansson (BCJ), this relation also holds in the field theory limit, giving rise to a very useful tool for computing quantum gravity amplitudes by ``squaring'' the amplitudes of gauge theories defined in the same number of dimensions \cite{Bern:2008qj,Bern:2010yg,Bern:2010ue,Bern:2010tq,Carrasco:2011mn,Oxburgh:2012zr,Bern:2013yya}.

A natural question is what the implications of the BCJ relations are for classical solutions. By ``squaring'' the numerator of some diagrams in Yang-Mills theory, it is possible to construct perturbative classical solutions of gravity theories \cite{Luna:2016hge}. On the other hand, for certain algebraically special spacetimes\footnote{Recently, there appeared a metric formulation of the classical double copy, also applicable to algebraically general spacetimes \cite{Kent:2025pvu}}, the non-linear nature of Einstein's equations can be overcome and it becomes possible to map these exact solutions to those of Maxwell's theory of electromagnetism. This research program, also pursued in the present work, is named the classical double copy (CDC).

There exist two methods to achieve linearity on the gravitational side. One is based on the observation that the Ricci tensor with mixed indices is linear in the perturbation for spacetimes admitting the Kerr-Schild (KS) coordinates \cite{Gurses:1975vu} (see \cite{Taub:1981evj,Xant1,Xant2} for various generalizations). Using this in trace-reversed Einstein's equations, one can obtain a map to solutions of Maxwell's theory, the so-called single copies, by making appropriate identifications \cite{Monteiro:2014cda}. This version of the CDC is called the Kerr-Schild double copy (KSDC) since it is restricted to spacetimes whose metrics can be written in the KS coordinates. The second method, the Weyl double copy (WDC), is based on spinorial techniques \cite{Luna:2018dpt}. For certain algebraically special spacetimes, one can find a spinor basis that linearizes the Weyl spinor and then relate it to the field strength spinor of a solution of Maxwell's theory\footnote{Solution of the $0$-th component of Maxwell's equation, i.e. Poisson's equation, also enters the formula. See subsection \ref{subsec:SWDC} for details.}. For vacuum solutions of general relativity (GR) admitting the KS coordinates, the two methods agree.

Although the WDC was shown to be valid for spacetimes not admitting the KS coordinates, and therefore more general in this sense, its original formulation in \cite{Luna:2018dpt} is restricted to vacuum solutions. Incorporation of the sources on the gravity side was achieved in \cite{Easson:2021asd,Easson:2022zoh} for type D solutions of Einstein-Maxwell theory. The key observation is that each term in the metric function should be considered separately. While the leading term is mapped to the  vacuum solution of Maxwell's theory, other terms are mapped to sourced solutions of Maxwell's theory one by one. The Weyl spinor of the gravitational solution is written as a sum of the expressions containing the product of field strength spinors corresponding to sourced solutions of Maxwell's theory.

In the KSDC, the spacetime metric is written as the sum of the metric of a background spacetime and a perturbation with special properties. In \cite{Monteiro:2014cda}, the background spacetime was taken to be flat. Motivated by the realization of the BCJ relations in certain curved backgrounds \cite{Adamo:2017nia}, the KSDC was extended to spacetimes with maximally symmetric backgrounds in \cite{CarrilloGonzalez:2017iyj}. AdS black hole solutions which are used to probe the properties of strongly coupled CFTs at finite temperatures through the AdS/CFT correspondence were studied as examples. This construction was extended to spacetimes with a general curved background in \cite{Alkac:2021bav}. In this setup, the gravitational solutions are mapped to solutions of Maxwell's theory defined on the curved background spacetime (see \cite{BahjatAbbas:2017htu} for an alternative formulation where the background spacetime is also copied). This most general formulation of the KSDC was applied to various Lifshitz black hole solutions. Because of the anisotropic scaling of the time coordinate, they can be used in non-relativistic extensions of the AdS/CFT correspondence \cite{Taylor:2008tg}. The metric of Lifshitz black holes can be written in the KS coordinates with the Lifshitz background \cite{Ayon-Beato:2014wla}, which is not maximally symmetric. In \cite{Alkac:2021bav}, single copies of various Lifshitz black hole solutions were obtained as examples. 

One might wonder whether the WDC can also be extended such that it matches these more general results of the KS version. In the AdS/CFT correspondence and its non-relativistic extensions, the dual CFTs are defined on the conformal boundary of the gravitational solutions. In light of the success of the WDC in capturing key aspects of the asymptotic structure of spacetimes \cite{Godazgar:2021iae,Adamo:2021dfg,Mao:2023yle}, such an extension could provide valuable insight into a possible connection between the CDC and holographic dualities. Although this is rather speculative, it is always an interesting exercise to check whether the spinor formulation of the CDC has any limitations.

A potential problem arises already in one of the simplest examples, i.e. the Schwarzschild-AdS$_4$ black hole. When written in the KS coordinates with a flat background metric, the effect of the cosmological constant on the single copy solution is a radial linearly increasing electric field sourced by a constant charge density filling all space \cite{Alkac:2021bav}. However, the contribution of it to the spacetime metric is conformally flat when considered separately. Because the Weyl tensor transforms uniformly under conformal transformations of the metric, it vanishes for conformally flat metrics. Therefore, the Weyl spinor vanishes for this term in the metric function and the SWDC seems to break down in this example\footnote{When the metric is written in the KS coordinates with an AdS$_4$ background, the contribution to the KS single copy also disappears and the SWDC works without any problem.} \cite{Alkac:2023glx}.

One has a much more complicated scenario for Lifshitz black holes. In addition to possible terms in the metric function for which the Weyl spinor vanishes because of conformal flatness, one might also have terms producing zero electric field on the gauge theory side, which is also a problem for the consistency of the SWDC with the KSDC (see subsection \ref{subsec:SWDC} for details). Therefore, the study of Lifshitz black holes is an important consistency check for the two different formulations of the CDC. In \cite{Alkac:2023glx}, it was shown by considering three different examples from the literature that the consistency can be achieved by employing a regularization procedure. Also, it was observed that since the Lifshitz black holes cannot be obtained as vacuum solutions of GR, there are terms in the Weyl spinor that are not related to the single copy gauge field. In this paper, we will consider three examples that have not been investigated yet with the aim of putting the SWDC for Lifshitz black holes to a more serious test. The novel features of the solutions that we will study are as follows:
\begin{enumerate}[I.]
	\item It is possible to have a case where two terms in the metric function that require regularization. As an example, we will consider the solution given in \cite{Brynjolfsson:2009ct}.
	\item Lifshitz black holes are also solutions of some higher-curvature gravity theories. As a prototype, we examine the solution in \cite{Cai:2009ac}, which results from an $R^2$-correction to the Einstein-Hilbert action, by considering the effect of this term on the field equations as an energy-momentum tensor resulting from matter fields.
	\item In addition to the Lifshitz black hole solutions with a planar horizon that are typically used in holographic applications, one might also find solutions with hyperbolic or spherical horizons. In \cite{Alkac:2023glx} and also in the present paper, static Lifshitz solutions with different horizon topologies are covered. However, when the event horizon is planar, by a coordinate transformation that is well-defined locally but not globally, one can obtain a stationary Lifshitz black hole solution. We will study the solution found in \cite{Herrera-Aguilar:2021top}. 
\end{enumerate}

The outline of this paper is as follows: In section \ref{sec:CDC}, after introducing Lifshitz spacetimes and static Lifshitz black holes, we review the formulations of the CDC that we will apply to Lifshitz black hole solutions. For the reader's convenience, we will collect all the formulae required for the examples studied later. Section \ref{sec:ex} presents a detailed study of three different examples whose importance is explained above. We give a summary of our results in section \ref{sec:sum}.

\section{Classical double copy for static Lifshitz black holes}\label{sec:CDC}

\subsection{Static Lifshitz black holes}
Before we discuss the classical double copy for static Lifshitz black holes, we first present the initial motivation for studying them in the context of holography, and also some of their important properties. Perhaps, the best starting point is to introduce Lifshitz field theories, which are non-relativistic field theories with the following Lifshitz symmetry group  \cite{Taylor:2015glc},
\begin{equation}
\begin{array}{ll}
H: & t \rightarrow t^{\prime}=t+a, \\
P^i: & x^i \rightarrow x^{\prime i}=x^i+a^i, \\
L^{i j}: & x^i \rightarrow  x^{\prime i}=L^i{ }_j x^j, \quad L^i{ }_j \in \mathrm{SO}(d),\\
\mathcal{D}_z: & t \rightarrow t^{\prime}=\lambda^z t ; \quad x^i \rightarrow x^{\prime i}=\lambda x^i.
\end{array}\label{Lifd}
\end{equation}
denoted by $\mathrm{Lif}_d(z)$. Here, $H$ and $P^i$ are translations of the time coordinate $t$ and the spatial coordinates $x^i$ ($i=1, \cdots, d$) while $L^{i j}$ are the spatial rotations. $\mathcal{D}_z$ is the anisotropic scale symmetry with the dynamical exponent $z$. Note that $z \neq 1$ introduces an anisotropic scaling of the time coordinate $t$ and generalizes the scale symmetry $\cD: x^\mu \rightarrow x^{\prime \mu}=\lambda x^\mu$ of relativistic CFTs. In condensed matter systems, there appear phase transitions with fixed points exhibiting such a scaling and Lifshitz field theories are useful toy models to study such phenomena \cite{Grinstein:1981rbe, Hornreich:1975zz}. 

In order to study these $(d+1)$-dimensional Lifshitz field theories holographically, one has to first construct a geometry realizing the same symmetries in ($d+2$)-dimensions and realize it as a solution to a gravitational theory, which was first achieved in \cite{Kachru:2008yh}. The Lifshitz spacetime with the following line element
\begin{equation}
\dd{s}^{2} =\ell^{2}\left[-r^{2 z}  \dd{t}^{2}+\frac{\dd{r}^{2}}{r^{2}}+r^{2} \dd \mathbf{x}^2 \right], \qquad 	\dd \mathbf{x}^2=\dd x_1^2+\ldots+\dd x_d^2,\label{Lifpre}
\end{equation}
where $\ell$ is a constant of length dimension, respects the $\mathrm{Lif}_d(z)$ symmetries in \eqref{Lifd} if the radial coordinate $r$ transforms as $ r \rightarrow r^{\prime} = \frac{r}{\l}$ under the scale transformation. Note that $z=1$ corresponds to the usual $\mathrm{AdS}_{d+2}$ geometry with $\mathrm{SO}(2, d+1)$ symmetry. One can easily show that the strong and the null energy conditions are satisfied for $z \geq 1$, which means that this geometry can be supported by physically reasonable matter provided that $z \geq 1$  \cite{Taylor:2015glc} (see \cite{Hoyos:2010at} for a discussion of pathologies arising when $z<1$).

On the other hand, a holographic description of the Lifshitz field theories at finite temperature requires a black hole solution that asymptotes to the Lifshitz spacetime with the line element \eqref{Lifpre}, which is called a Lifshitz black hole. Its line element can be taken as
\begin{equation}
\dd{s}^{2} =\ell^{2}\left[-h(r) r^{2 z}  \dd{t}^{2}+\frac{\dd{r}^{2}}{h(r) r^{2}}+r^{2} \dd \mathbf{x}^2 \right],\label{Lifbhpre}
\end{equation}
where $h(r)$ is the metric function satisfying $h \to 1$ as $r \to \infty$, which ensures that we have an asymptotically Lifshitz black hole. It should also have at least one zero [$h(\rp) = 0$] such that we can define the temperature of the black hole thanks to the event horizon located at $r = \rp$, and identify it with the temperature of the dual Lifshitz field theory. Once we find a matter coupling (or a higher-curvature modification) that admits a Lifshitz black hole, the properties of the finite-temperature dual field theory at the strong coupling can be studied holographically within the gravitational theory defined in one higher dimension. We would like to emphasize that there is no systematic way to find a theory admitting Lifshitz black holes, but fortunately a reasonable amount of examples exist.

So far, we have assumed that the dual field theory is defined on $(d+1)$-dimensional Minkowski spacetime with $1$ time and $d$ spatial coordinates, denoted by $\mathrm{R}^{1,d}$. However, for the spatial coordinates, instead of a planar geometry $\mathrm{R}^d$, we can also assume a spherical ($\mathrm{S}^d$) or a hyperbolic ($\mathrm{H}^d$) geometry, and define the corresponding dual field theories on $\mathrm{R}^1 \times \mathrm{S}^d$ and $\mathrm{R}^1 \times \mathrm{H}^d$ respectively. For a finite-temperature holographic description, the line element in \eqref{Lifbhpre} should be generalized as follows
\begin{equation}
\dd{s}^{2} =\ell^{2}\left[-r^{2 z}  h(r)  \dd{t}^{2}+\frac{\dd{r}^{2}}{r^{2}h(r)}+r^{2} \dd \Sigma_d^2 \right],\label{Lifbhpre2}
\end{equation}
where $\dd \Sigma_d^2$ is the line element of $\mathrm{S}^d$, $\mathrm{R}^d$ or $\mathrm{H}^d$, corresponding to a black hole with a spherical, planar or hyperbolic event horizon respectively. In the next subsection, we will give an appropriate parametrization for $d=2$ and study Lifshitz black holes in $D=4$.

\subsection{Kerr-Schild double copy}\label{subsec:KSDC}
As mentioned in the introduction, the KSDC is formulated starting from the trace-reversed Einstein's equations with mixed indices given by
\begin{equation}
	R^\m_{\  \n} -\frac{2\,\Lambda}{D-2}\, \d^\m_{\  \n}=\tT^\m_{\  \n},\label{reversed}
\end{equation}
where $\L$ is the cosmological constant, $D$ is the number of spacetime dimensions, and $\widetilde{T}^\m_{\  \n}$ is the trace-reversed energy momentum tensor defined by 
\begin{equation}
	\tT^\m_{\  \n} = T^\m_{\  \n} -\frac{1}{D-2}\,\d^\m_{\  \n}\,T, \qquad T = T^\m_{\  \m}.
\end{equation}
We will follow the treatment of \cite{Alkac:2021bav} where no simplifying assumption about the background metric is made. We assume that the spacetime admits the KS coordinates and the metric can be written in the following form
\begin{equation}
	g_{\m\n}=\bar{g}_{\m\n}+\phi\, k_\m k_\n,\label{KS}
\end{equation}
where $\bar{g}_{\m\n}$ is the background metric. $\f$ is the KS scalar and the vector $k_\m$ is null and geodesic with respect to the background and the full metric. Thanks to this special form of the perturbation, the Ricci tensor with mixed indices becomes \cite{Stephani:2003tm}
\begin{equation}
	R^{\mu}_{\ \nu}=\bar{R}^{\mu}_{\ \nu}-\phi\, k^{\mu} k^{\a} \bar{R}_{\a \nu}+\frac{1}{2}\left[\bar{\nabla}^{\a} \bar{\nabla}^{\mu}\left(\phi \,k_{\a} k_{\nu}\right)+\bar{\nabla}^{\a} \bar{\nabla}_{\nu}\left(\phi\, k^{\mu} k_{\a}\right)-\bar{\nabla}^{2}\left(\phi \,k^{\mu} k_{\nu}\right)\right]\label{ricci},
\end{equation}
which is linear in the perturbation. We make the identification \cite{Monteiro:2014cda}
\begin{equation}
	A_\m \equiv \f\, k_\m,\label{Adef}
\end{equation}
where $A_\m$ is the single copy gauge field and write the Ricci tensor in \eqref{ricci} as
\begin{equation}
	R^{\m}_{\ \n}= {\bR^\mu}_{\ \n}-\frac{1}{2}\left[ \bar{\nabla}_{\a} F^{\a \mu}k_\n+	E^{\m}_{\ \n}\right].\label{riccif}
\end{equation}
Here, $F_{\m\n}=2\,\bnab_{[\m}A_{\n]}$ is the field strength tensor and $E^{\m}_{\ \n}$ is a tensor of the following complicated form
\begin{align}
	E^{\m}_{\ \n}=& -\bar{\nabla}_{\nu}\left[A^{\mu}\left(\bar{\nabla}_{\a} k^{\a}+\frac{k^{\a} \bar{\nabla}_{\a} \phi}{\phi}\right)\right] + F^{\a \mu} \bar{\nabla}_{\a} k_{\nu}-\bar{\nabla}_{\a}\left(A^{\a} \bar{\nabla}^{\mu} k_{\nu}-A^{\mu} \bar{\nabla}^{\a} k_{\nu}\right) \nn \\
	&-\bar{R}_{\ \a \b \n}^{\m} A^{\a} k^{\b} +\bar{R}_{\alpha \n} A^{\alpha} k^{\mu}.
\end{align}
Next we use the expression for the Ricci tensor \eqref{riccif} in the trace-reversed Einstein's equations \eqref{reversed} and consider the contraction with a Killing vector $V^\n$ of both the full and the background metrics. After some manipulations (see \cite{Alkac:2021bav} for more details), we find the following equation for the single copy gauge field
\begin{equation}
	\bar{\nabla}_{\n} F^{\n \mu} + E^\m=J^\m,\label{single}
\end{equation}
where $E^\m$ is an ``extra'' contribution to Maxwell's equation defined on the background spacetime that takes the following form
\begin{equation}
E^\m=\frac{1}{V \cdot k}\,E^\m_{\ \n}\, V^\n. \label{Edef}
\end{equation}
The source is given by 
\begin{equation}
	J^\m = 2 \left[\D^\m-\tT^\m\right]\label{Jdef},
\end{equation}
\begin{equation}
	\D^\m = \frac{1}{V \cdot k}\, \D^\m_{\  \n} V^\n, \qquad \qquad \tT^\m = \frac{1}{V \cdot k}\, \tT^\m_{\  \n} V^\n.\label{deltadef}
\end{equation}
The second term is the expected contribution from the energy-momentum tensor. The first term becomes non-zero when the background spacetime deviates from a maximally symmetric spacetime whose (A)dS length is related to the cosmological constant as
\begin{equation}
	\L = \e \, \frac{(D-1)(D-2)}{2\,L^2},\label{adscosm}
\end{equation}
where $\e = -1, +1$ corresponds to AdS and dS cases respectively. The tensor $\D^\m_{\ \n}$ that measures the deviation from the maximally symmetric spacetime is given by
\begin{equation}
	\D^\m_{\  \n} =\bR^\m_{\  \n}-\frac{2\,\L}{D-2}\, \d^\m_{\ \n}.\label{delta} 
\end{equation}
Contracting \eqref{single} with the Killing vector $V^\m$ once more, we obtain the equation for the zeroth copy field $\f$ as
\begin{eqnarray}
\bar{\nabla}^2 \phi + \mathcal{C} +\mathcal{E}  = j,\label{zeroth}
\end{eqnarray}
where
\begin{equation}
\mathcal{C} = \frac{V \cdot C}{V \cdot k},\qquad \qquad \qquad \mathcal{E} = \frac{V \cdot E}{V \cdot k}, \qquad \qquad \qquad j=\frac{V \cdot J}{V \cdot k}.\label{defs}
\end{equation}
The vectors $E^\m$ and $J^\m$ are already given in (\ref{Edef}, \ref{Jdef}) and,
\begin{equation}
C^\m=\bar{\nabla}_\a k^\m\, \bar{\nabla}^\a \phi + \bar{\nabla}_\a \left[2 \phi \bar{\nabla}^{ [ \a} k^{ \m] } - k^\a \bar{\nabla}^{\m} \phi\right].\label{Zdef}
\end{equation}
The $j$-term and the $\cC$-term are the source term and the usual curvature modification of the Poisson's equation respectively. The $\cE$-term is the remnant of the ``extra'' term $E^\m$ that first appeared in \eqref{single}. 

One might say it can be just embedded into the definition of the source in \eqref{Jdef}. However, it contains only terms related to the metric perturbation, and therefore, does not have a natural interpretation as a source unlike the other terms in the definition. Without considering any specific spacetime admitting the KS coordinates that is a solution to the trace-reversed Einstein's equations \eqref{reversed}, the equations (\ref{single}, \ref{zeroth}) are the simplest form of the single and zeroth copy equations. However, it is quite remarkable that the extra term $E^\m$ lacking a natural interpretation and its remnant $\cE$ vanish for all the examples studied so far including the Lifshitz black holes. It would be interesting to have a general proof, however, this seems unlikely since there is no systematic method for writing a general metric in the KS coordinates and the explicit forms of the KS scalar $\f$ and the vector $k_\m$ are important.

After this general discussion, we will now consider Lifshitz black holes in $D=4$, which allows us to make a comparison with the WDC since the spinorial techniques are available in this spacetime dimension. Using the general form in \eqref{Lifbhpre2}, we take the following line element
\begin{equation}
\dd{s}^{2}=\ell^{2}\left[-r^{2 z} h(r) \dd{t}^{2}+\frac{\dd{r}^{2}}{r^{2} h(r)}+r^{2} \left[\dd^2{\th} + \chi^2(\th) \dd^2{\f}\right]\right].\label{lifbh}
\end{equation}
The event horizon topology is parameterized as
\begin{equation}
\chi(\theta)=\left\{\begin{array}{ccl}
\sin \theta & \text { if } & k=1, \\
\theta & \text { if } & k=0, \\
\sinh \theta & \text { if } & k=-1,
\end{array}\right.
\end{equation}
where $k = 1, 0, -1$ correspond to a spherical, planar and hyperbolic cases. It is possible to obtain compact sections at constant $t$ and $r$ through appropriate identifications \cite{Smith:1997wx, Mann:1997iz}.

Our task now is to write the metric in the KS coordinates. By applying the following coordinate transformation introduced in \cite{Ayon-Beato:2014wla}
\begin{eqnarray}
\dd{t} \rightarrow \dd{t} +\frac{(h-1) r^{-(z+1)}}{h} \,\dd{r},\label{trans} 
\end{eqnarray}
to the line element \eqref{lifbh}, we obtain the metric in the KS coordinates with the background metric 
\begin{equation}
\dd{\bar{s}}^{2}=\ell^{2}\left[-r^{2 z}  \dd{t}^{2}+\frac{\dd{r}^{2}}{r^{2}}+r^{2}\left[\dd^2{\th} + \chi^2(\th) \dd^2{\f}\right]\right].\label{lif}
\end{equation}
This is a generalization of the Lifshitz spacetime with the line element \eqref{Lifpre} in $D=4$, such that it is the asymptote of the black holes with non-trivial event horizon topologies, which have the line element \eqref{lifbh} (since $h \to 1$ as $r \to \infty$). For simplicity, we will continue to refer to this geometry as the Lifshitz spacetime, as the precise meaning should be clear from the context.

With this background metric, the KS scalar and the null vector read
\begin{equation}
\phi=\ell^{2}\,(1-h)r^{2z}, \qquad \qquad k_{\mu} \dd{x}^{\mu}= \dd{t}+{\frac{\dd{r}}{r^{z+1}}}.\label{kandf}
\end{equation}
For black hole solutions, we use the time-like Killing vector\footnote{As shown in \cite{CarrilloGonzalez:2017iyj}, one can also study the gauge theory counterpart of wave-type solutions (of Petrov type N) by choosing a null Killing vector.} $V^\m = \d^\m_{\ t}$, which leads to
\begin{equation}
V \cdot k = 1, \qquad E^\m=E^{\m}_{\ 0} = 0, \qquad \cE=E^0=0, \qquad \D^\m = \D^\m_{\  0}, \qquad \tT^\m = \tT^\m_{\  0}.
\end{equation}
Remarkably, for a generic KS scalar $\f$, the ``extra'' term $E^\m$ in \eqref{single} vanishes. This is quite non-trivial since the tensor $E^\m_{\ \n}$ does not vanish but the time-like Killing vector $V^\n$ is always an eigenvector of it with a zero eigenvalue, ensuring $E^\m = 0$. As a result, we get the Maxwell's and Poisson's equations for the single copy field $A_\m$ and the zeroth copy field $\f$ defined on the Lifshitz background spacetime given in \eqref{lif} as follows
\begin{align}
\bar{\nabla}_{\n} F^{\n \mu} &=J^\m,\label{maxfin} \\
\bar{\nabla}^2 \phi + \cC &= j, \label{poisfin}
\end{align}
where the source $J^\m$ is defined in \eqref{Jdef} and $j=J_0$. When the event horizon is a plane ($k=0$), the curvature modification $\cC$ to the Poisson's equation \eqref{zeroth} takes a particularly simple form and the zeroth copy equation can be written as
\begin{equation}
	\bar{\nabla}^2 \phi + \frac{(z-2) z}{z^2+2 z+3} \bar{R} \f  = j,
\end{equation}
which is a generalization of the Poisson's equation on the AdS spacetime corresponding to $z=1$.

This concludes our review of the KSDC for static Lifshitz black holes. When we study a stationary Lifshitz black hole in subsection \ref{subsec:sta}, we will discuss the necessary modifications in the procedure.

\subsection{Sourced Weyl double copy}\label{subsec:SWDC}
For the spinorial formalism of GR, we mainly use \cite{Penrose:1960eq}. The excellent summary in the appendix of \cite{Easson:2022zoh} is enough to follow the discussion here and to reproduce our results. We adopt the same conventions with this work. For pedagogical introductions, the reader is referred to \cite{ODonnell:2003lqh,Penrose:1985bww,Penrose:1986ca}. We find the concise treatment of \cite{ODonnell:2003lqh} particularly useful. 

Following \cite{Luna:2018dpt}, it is easy to understand the WDC for vacuum solutions of GR. Any Weyl spinor can be decomposed into four rank-1 spinors as follows
\begin{equation}
\Psi_{A B C D}=\alpha_{(A} \beta_B \gamma_C \delta_{D)},
\end{equation}
where these four spinors are the four principal null directions of the spacetime. The classification of a spacetime is done based on how many of these null directions coincide up to scaling. If we have four distinct principal null directions, the spacetime is algebraically general and called of type I. The two Petrov types discussed in \cite{Luna:2018dpt} are type D and type N spacetimes, which correspond to possessing two principal null directions with multiplicity two and a single principal null direction with multiplicity four respectively. For type D spacetimes, which we are interested in here, one has
\begin{equation}
\Psi^\text{[type D]}_{A B C D}=\alpha_{(A} \beta_B \alpha_C \beta_{D)}.
\end{equation}
Since it is possible to map any anti-symmetric tensor into a 2-spinor, one can also classify the field strength similarly. In general, we have the following decomposition
\begin{equation}
	f_{AB} = \x_{(A} \z_{B)},
\end{equation}
with two possible principal null directions. When they are distinct, the field strength spinor is algebraically general. It is algebraically special when they coincide. This tells us we might expect a relation of the following form
\begin{equation}
	\Psi^\text{[type D]}_{A B C D} \sim f^\text{[gen]}_{(AB} f^\text{[gen]}_{CD)},
\end{equation}
where $f^\text{[gen]}_{AB}$ is an algebraically general field strength. By the same logic, it is also possible to have the relation $\Psi^\text{[type N]}_{A B C D} \sim f^\text{[sp]}_{(AB} f^\text{[sp]}_{CD)}$ between the Weyl spinor of type N spacetimes and the algebraically special field strengths $f^\text{[sp]}_{AB}$ (see \cite{Godazgar:2020zbv} for the WDC of radiative solutions of the vacuum Einstein's equations).

For type D spacetimes, this expectation from the algebraic classification of Weyl and the field strength spinors can be realized as a theorem \cite{Walker:1970un, Hughston:1972qf, Dietz361} that also fixes the ``proportionality factor''. Considering the Weyl spinor of a type D vacuum solution admitting a rank-2 Killing spinor, we have
\begin{equation}
	\Psi_{A B C D} =\frac{1}{S} f_{(AB} f_{CD)},\label{weylvac}
\end{equation}
where $f_{AB}$ is the field strength spinor of the solution of Maxwell's equation on the curved background with the Weyl spinor $\Psi_{A B C D}$. The scalar field $S$ satisfies the Poisson's equation on this curved background (see \cite{Luna:2018dpt} for explicit expressions in terms of the Killing spinor). However, when the spacetime admits the KS coordinates, the Maxwell's and the Poisson's equations are identical for the full and the background spacetime because of the property $\det g = \det \bar{g}$. The scalar $S$, which is determined by the norm of the Killing spinor, is in general complex. However, a linear combination of its real and imaginary parts should be related to the zeroth copy, and as a result, the right-hand-side of the equation \eqref{weylvac} also follows from the single and the zeroth copies defined on the background spacetime. This gives us the spinorial form of the CDC, the WDC.

Studying more general solutions, it was observed that the following modifications are needed in the WDC to yield the spinorial version of the results obtained in the KSDC:
\begin{enumerate}
	\item When sources are present on the gravity side, one should consider each term in the metric function separately and one has a sum of scalar-gauge theory solutions at the right hand side of \eqref{weylvac} \cite{Easson:2021asd}. The leading term comes from the vacuum solution if the metric has such a piece.
	\item When the gravitational solution has no vacuum piece, there appear terms in the Weyl tensor that are irrelevant to the properties of the single and the zeroth copies \cite{Alkac:2023glx}. Such terms should be neglected in the WDC since this is just a consequence of the fact that solutions with certain symmetries can only be obtained by matter coupling.
	\item The conformally flat pieces of the metric function do not lead to a non-zero Weyl spinor, which is required for the agreement with the KSDC. It is also possible to get a vanishing contribution to the field strength spinor. Such terms should be dealt with the regularization procedure of \cite{Alkac:2023glx}.
\end{enumerate}
As a result, the form of the SWDC that was shown to be valid for the examples considered so far is as follows
\begin{equation}
\Psi_{A B C D} = \ldots +  \sum_{i=1}^N \frac{1}{S_{(i)}}f^{(i)}_{(A B}f^{(i)}_{C D)},\label{SWDC}
\end{equation}
where $\ldots$ denotes the above-mentioned irrelevant terms and, whenever necessary, the regularization of \cite{Alkac:2023glx} should be employed for the terms in the sum.

In a suitable spinor basis $\{o_A, \iota_B\}$, the Weyl spinor for type D spacetimes can be written as
\begin{equation}
	\Psi_{A B C D} = 6 o_{(A} \iota_B o_C \iota_{D)}  \Psi_{2},\label{PsiD}
\end{equation}
where $\Psi_2$ is the only-nonvanishing Weyl scalar for this type of spacetimes (see \cite{Easson:2022zoh} for the calculation of the Weyl scalars). For static black holes, the single copy field strength spinor take the following form in the same basis
\begin{equation}
	f_{A B} = Z\, o_{(A} \iota_{B)},\label{fbasis}
\end{equation}
where $Z$ is real and can be found in terms of the radial single copy electric field
\begin{equation}
	E \equiv F_{rt},
\end{equation}
which is the only non-zero independent component of the field strength tensor. The field $S$ in \eqref{SWDC} becomes identical with the zeroth copy $\f$ and one can write down the SWDC \eqref{SWDC} as a relation between the Weyl scalar $\Psi_2$, the scalar $Z$ and the zeroth copy $\f$ and one finds the following \emph{consistency condition} 
\begin{equation}
\Psi^{(i)}_{2} \propto \frac{Z_{(i)}^2}{\phi_{(i)}},\label{CC}
\end{equation}
which needs to be satisfied for matching the results in the KSDC up to irrelevant terms denoted by $\ldots$.

Modifications 2 and 3 were obtained by studying static Lifshitz black holes. In order to introduce the regularization procedure, let us consider the line element \eqref{lifbh}  of a general static Lifshitz black hole and take a metric function of the form
\begin{equation}
	h = 1 + \sum_{n} \frac{a_n}{r^n}.\label{h}
\end{equation}
Applying the general formulation of the KSDC described in the previous sub-section, one finds \cite{Alkac:2021bav}
\begin{equation}
	\f = -\ell^{2} \sum_{n} \frac{a_n}{r^{n-2z}}, \qq E =  \ell^{2} \sum_{n} \frac{(n-2z) a_n}{r^{n-2z+1}}.\label{fande}
\end{equation}
The Weyl scalar reads
\begin{equation}
	\Psi_2 = \frac{z(z-1)}{6 \ell^{2}} - \frac{ k}{6 \ell^{2} r^2}  + \sum_{n} \frac{(n-z)(n-2z+2)\, a_n }{12\ell^{2}r^n},\label{psilif}
\end{equation}
and the scalar $Z$ satisfies
\begin{equation}
	Z \propto r^{1-z} E,\label{Zlif}
\end{equation}
where the irrelevant numerical factors can be omitted \cite{Alkac:2023glx}. From equations (\ref{fande}, \ref{psilif}), we observe that there are certain critical values of $n$, denoted by $n_*$, for which the scalar $Z$ or the Weyl scalar $\Psi_2$ vanishes, and as a result the consistency condition \eqref{CC} cannot be satisfied. The critical values are $n_* = z, 2z - 2$, which make $\Psi_2 = 0$ \footnote{When $z=2$, the two critical values that make $\Psi_2 = 0$ coincide. An example of this was studied in \cite{Alkac:2023glx}.}, and $n_* = 2z$, which gives $Z=0$ [through the vanishing of the electric field $e$ in \eqref{fande}].

In \cite{Alkac:2023glx}, it was shown that such terms can be regularized by calculating the vanishing terms in the consistency condition \eqref{CC} more carefully as follows
\begin{enumerate}
	\item Use an arbitrary exponent $n$ instead of the critical value $n_*$ in the relevant term as $\frac{a_{n_*}}{r^n}$.
	\item Scale the coefficient as $a_{n_*} \to \frac{a_{n_*}}{n-n_*}$ and calculate the vanishing term in the consistency condition ($\Psi_2$ or $Z$) with this coefficient.
	\item Insert the actual value of the exponent ($n = n_*$).
\end{enumerate}
In this way, one obtains a non-zero  $\Psi_2$ or $Z$ with which the SWDC becomes consistent with the KSDC.

In the next section, we will apply it to two static Lifshitz black hole solutions that have not been studied previously. Also, we will study the SWDC for a stationary Lifsthiz black hole solution for the first time in the literature.

\section{New examples}\label{sec:ex}
\subsection{Example I: Two terms are regularized}
Our first example is a charged generalization of the Lifshitz topological black holes \cite{Mann:2009yx}, found in \cite{Brynjolfsson:2009ct}. Let us consider an action of the form 
\begin{equation}
	S=\int \dd^4 x \sqrt{-g}\left( R - 2 \Lambda +\cL_m \right),\label{actgen}
\end{equation}
where the matter Lagrangian reads
\begin{equation}
	\cL_m = - \frac{1}{4}\mathcal{F}_{\m\n} \mathcal{F}^{\m\n} - \frac{1}{12} \cH_{\m\n\r} \cH^{\m\n\r} - C \epsilon^{\m\n\r\s} \mathcal{B}_{\m\n}\mathcal{F}_{\r\s} - \frac{1}{4} \mathcal{G}_{\m\n} \mathcal{G}^{\m\n}.\label{ex1mat}
\end{equation}
$\mathcal{F}_{\m\n}$ is a two-form, $\cH_{\m\n\s} = 3 \partial_{[\m}$ $\mathcal{B}_{\n\s]}$ is a three-form and they are topologically with the coupling constant $C$. The two-form $\cG_{\m\n}$ gives rise to a charged version of the Lifshitz topological black holes that we study here.

With the matter coupling given in \eqref{ex1mat}, one obtains the field equations \eqref{reversed} with the following trace-reversed energy-momentum tensor
\begin{equation}
	\tT^\m_{\ \n} = \frac{1}{4} \left[2 \cF^{\m\a} \cF_{\n\a} -\frac{1}{2} \d^\m_{\ \n} \cF^2 + 2 \cG^{\m\a} \cG_{\n\a} -\frac{1}{2} \d^\m_{\ \n} \cG^2 + \cH^{\m\a\b} \cH_{\n \a \b} -\frac{1}{2} \d^\m_{\ \n} \cH^2\right], 
\end{equation}
where we used $\cF^2 = \cF^{\m\n} \cF_{\m\n}$, $\cG^2 = \cG^{\m\n} \cG_{\m\n}$, and $\cH^2 = \cH^{\m\n\r} \cH_{\m\n\r}$. The matter fields satisfy the following equations
\begin{equation}
	\nabla^\a \cF_{\mu \a}  =-\frac{C}{6} \epsilon_{\mu \nu \alpha \beta} \cH^{\nu \alpha \beta}, \qq \nabla^\a \cH_{\mu \nu \a} =\frac{C}{2} \epsilon_{\mu \nu \alpha \beta} F^{\alpha \beta}, \qq \nabla^\a \cG_{\mu \a}=0.
\end{equation}
We will consider static Lifshitz black hole solutions with spherical, planar and hyperbolic event horizon topologies. Therefore, the machinery summarized in section \ref{sec:sum} is directly applicable.

A solution with a line element of the form \eqref{lifbh} was given in \cite{Brynjolfsson:2009ct}, which has the following metric function and the Lifshitz exponent
\begin{equation}
	h = 1+\frac{ k }{10 r^2}-\frac{ 3 k^2}{400 r^4}-\frac{q^2}{2 \ell^2 r^4},\qq z=4,\label{ex1met}
\end{equation}
where $q$ is the electric charge. This solution is supported by the following matter configuration
\begin{equation}
	\mathcal{F}_{rt} = - \frac{\sqrt{6} \ell }{10}  \left(20 r^3+k r\right),\qquad \cH_{r\theta\phi} = 2 \ell^2 \sqrt{3  \chi(\theta)} r,\qquad \mathcal{G}_{rt}=2 \ell q r,\label{ex1matconf}
\end{equation}
provided that the cosmological constant $\L$ and the coupling constant $C$ is chosen as
\begin{equation}
	\Lambda = \frac{12}{\ell^2},\qq \qq C = \pm \frac{2 \sqrt{2}}{\ell}.
\end{equation}
Note that they take the same form both in the Boyer-Lindquist coordinates [used in \eqref{lifbh}] and in the KS coordinates.

From our general analysis, we know that the critical exponents in the metric function is $n_*= 4, 6, 8$ for $z=4$. The first two values give a vanishing Weyl scalar $\Psi_2$, the last one gives a zero electric field $E$, and a zero scalar $Z$. This implies that the last two terms in the metric function \eqref{ex1met} do not contribute to the Weyl scalar. Note that they have a different origin. While the former is a part of the uncharged solution, the latter arises after introducing an electric charge $q$ through the two-form $\cG_{\m\n}$. 

Let us see explicitly how they violate the consistency condition \eqref{CC}. The KS scalar and the electric field can be read from \eqref{fande} as
\begin{align}
		\f & =  \frac{1}{400} r^4 \left[k  \ell^2 \left(3 k -40 r^2\right)+200 q^2\right] \\
		E & = \left[- \frac{3 k \ell^2}{5}r^2 +\frac{3 k^2 \ell^2}{100}   +2 q^2\right]r^3.
\end{align}
The sources of the single and the zeroth copy equations (\ref{maxfin}, \ref{poisfin}) are given by
\begin{equation}
	J^\m \pd_\m =  \frac{3 k  \ell^2 \left(40 r^2-k \right)-200 q^2}{50 \ell^4 r^4} \pd_t, \qq j = \frac{4 q^2 r^4}{\ell^2}+\frac{3}{50} k  r^4 \left(k -40 r^2\right)
\end{equation}
As expected, each term in the metric function gives a non-zero contribution to the KS scalar and the electric field. Calculating the scalars $\Psi_2$ and $Z$ from (\ref{psilif}, \ref{Zlif}), we find
\begin{align}
	\Psi_2 &= \frac{2}{\ell^2} -\frac{k }{6 \ell^2 r^2} + \frac{ k}{15  r^2},\\
	Z  & \propto \left[-\frac{3k  }{5} r^2+\frac{3 k^2}{100}+\frac{2 q^2}{\ell^2}\right]r^3.
\end{align}
There are three different contributions to the scalar $Z$. However, the first two term in the Weyl scalar $\Psi_2$ are the terms that are irrelevant for the single copies whose general form is given in \eqref{psilif}. The remaining term is the contribution of the $n=2$ term in the metric function and the two $n=4$ terms do not contribute due to the conformal flatness. In order to recover the missing terms, the regularization procedure described in the previous section should be applied as follows: Taking the $n=4$ terms in the metric function as $\frac{a_4}{r^n}$ with
\begin{equation}
		a_4 = -\frac{3 k^2}{400} - \frac{q^2}{2 \ell^4},\label{ex1a4}
\end{equation}
the Weyl scalar $\Psi_2$ becomes,
\begin{equation}
	\Psi_2 = \frac{2}{\ell^2} -\frac{k }{6 \ell^2 r^2}+ \frac{k}{15 r^2} + \frac{(n-4)(n-6) a_4}{12 \ell^2 r^4},
\end{equation}
which explicitly shows why the contribution from these terms vanish. By scaling the coefficient as $a_4 \to \frac{a_4}{n-4}$, and then setting $n=4$, we obtain the following regularized Weyl scalar
\begin{equation}
	 \Psi_2^{\text{[reg]}} = \frac{2}{\ell^2} -\frac{k }{6 \ell^2 r^2}+ \frac{k}{15 r^2} - \frac{a_4}{6 \ell^2 r^4},
\end{equation}
where $a_4$ is given in \eqref{ex1a4}. This regularized Weyl scalar satisfies the consistency condition \eqref{CC} term-by-term.

\subsection{Example II: Solution arising from an $R^2$-correction}
As mentioned in the introduction, the Lifshitz black holes are also solutions to some higher-curvature gravity theories. For an action of the form \eqref{actgen}, we can consider the effect of an $R^2$-correction by taking 
\begin{equation}
	\cL_m = \a R^2,
\end{equation}
which leads to the following trace-reversed energy-momentum tensor
\begin{equation}
	\widetilde{T}^\m_{\ \n} =  2 \alpha\left[2 \delta^\m_{\ \n} \nabla^2 R + \nabla^\mu \nabla_\nu R -  \left(R^\m_{\ \n} -\frac{1}{4} \delta^\m_{\ \n}  R\right) R \right].
\end{equation}
One might argue that the contributions from the pure gravitational terms in the action should be kept at the left hand side of the Einstein's equations and the contribution to the field equations should be evaluated with the KS ansatz. Such an approach was adopted in \cite{Alkac:2024pfd} and it was shown that one can establish a map between a class of gravity theories and Maxwell's theory where the single copy field is a polynomial in the KS scalar in $d>4$. The spinorial counterpart of this map is yet to be studied (see \cite{Zhao:2024ljb,Zhao:2024wtn} for some 5d results in GR). However, $f(R)$ gravities do not belong to that class. Here, we aim to show that regarding higher-curvature corrections in the action as effective matter fields also yields sensible results. In 3d KSDC, this idea was already applied in \cite{Alkac:2022tvc} to non-minimally coupled matter fields and the curvature terms were treated as usual matter source terms.

As an example, we will study the solution in \cite{Cai:2009ac}, which possesses a planar horizon ($k = 0$). The metric function and the Lifshitz exponent are as follows
\begin{equation}
	h= 1- \frac{r_+^3}{r^3},\qq \qq z=\frac{3}{2},
\end{equation}
where $r_+$ is the location of the event horizon. This is a solution if the coupling constant $\a$ and the cosmological constant $\L$ are fixed as follows
\begin{equation}
	\alpha =\frac{1}{33 \ell^2},\qq \qq \Lambda =-\frac{33}{8 \ell^2}.
\end{equation}
For the Lifshitz exponent $z=3/2$, the critical exponents are $n_* = 3/2, 1, 3$. In the metric function, we have the third critical exponent for which the scalar $Z$ becomes zero. The quantities relevant in the consistency condition \eqref{CC} follow from the general expressions (\ref{fande}-\ref{Zlif}) as
\begin{equation}
	\f = \frac{r_+^3}{\ell^3}, \qq Z \propto 0, \qq \Psi_{2} = \frac{1}{8 \ell^2} - \frac{r_+^3}{4 \ell^2 r^3}.	
\end{equation}
Remarkably, the Maxwell's \eqref{maxfin} and Poisson's equations \eqref{poisfin} are satisfied with vanishing sources and we have vacuum solutions. The first term in the Weyl scalar $\Psi_2$ is irrelevant. Obviously, the consistency condition cannot be satisfied with a vanishing $Z$, which can however be regularized by taking the relevant term in the metric function as $\frac{a_{3}}{r^n}$ with $a_3 = - r_+^3$. Now we have,
\begin{equation}
	Z = -\frac{(n-3) \ell^2 a_3}{r^{\nicefrac{7}{2}}}.
\end{equation}
After taking $a_3 \to \frac{a_3}{n-3}$, and then setting $n=3$, we obtain the following regularized $Z$ scalar
\begin{equation}
	Z^{\text{[reg]}} = -\frac{\ell^2 a_3}{r^{\nicefrac{7}{2}}},
\end{equation}
which now satisfies the consistency condition.

\subsection{Example III: Stationary Lifshitz black hole solution of [34]}\label{subsec:sta}
Now, we will study a stationary Lifshitz black hole arising from the Einstein-Maxwell-dilaton system with the matter Lagrangian
\begin{equation}
	\mathcal{L}_{m}=-\frac{1}{2} \nabla_\mu \vf \nabla^\mu \vf-\frac{1}{4} e^{-\lambda \vf} \cF_{\mu \nu} \cF^{\mu \nu},
\end{equation}
where $\l$ is a constant, $\vf$ is the dilaton field and $\cF_{\m\n}$ is the Maxwell two-form. The trace-reversed energy-momentum tensor can be found as 
\begin{equation}
	\tT^\m_{\ \n} =\frac{1}{2} \nabla^\mu \varphi \nabla_\nu \varphi+\frac{1}{2} e^{\lambda \varphi} \cF^{\mu \alpha} \cF_\nu^\alpha-\frac{1}{8} \d^\m_{\ \n} e^{\lambda \varphi} \cF_{\alpha \beta} \cF^{\alpha \beta},
\end{equation}
and the matter fields satisfy
\begin{align}
	 \partial_\mu\left(\sqrt{-g} e^{\lambda \varphi} \cF^{\mu \nu}\right)&=0 \\
	 \partial_\mu\left(\sqrt{-g} \partial^\mu \varphi\right)-\frac{\lambda}{4} \sqrt{-g} e^{\lambda \varphi} \cF_{\mu \nu} \cF^{\mu \nu}&=0 .
\end{align}
The static Lifshitz black hole solution of this theory found in \cite{Mann:2009yx} was studied in the context of CDC in \cite{Alkac:2021bav,Alkac:2023glx}. Remarkably, it is mapped to the vacuum solution of Maxwell's theory defined on the Lifshitz spacetime with the line element \eqref{lif}, which is matched in the SWDC after regularization when required. For our analysis, we will take the line element of a static black hole solution slightly different than \eqref{lifbh} to follow the conventions of \cite{Herrera-Aguilar:2021top} where the stationary version is given. 

Consider the following line element for a static Lifshitz black hole with a planar horizon topology ($k=0$)
\begin{equation}
\dd{s}^2 = -\frac{r^{2z}h(r)}{\ell^{2z}}  \dd{t}^2+ \frac{\ell^2 \dd{r}^2}{r^2 h(r)}  + r^2\dd{\th}^2 + \frac{r^2}{\ell^2} \dd{x}^2,
\end{equation}
where the $\th$-coordinate is compactified ($0 \leq \th < 2\pi$) and $- \infty < x < \infty$. For the following values of the cosmological constant $\L$ and the constant $\l$
\begin{equation}
	\Lambda = -\frac{(z+2)(z+1)}{2 \ell^2},\qq \quad \lambda=-\frac{2}{\sqrt{z-1}},
\end{equation}
the theory admits a static black hole solution with the following metric function and the Lifshitz exponent
\begin{equation}
	 h = 1- \left(\frac{r_+}{r}\right)^{z+2}, \qq z \geq 1.\label{ex3met}
\end{equation}
The matter configuration is as follows
\begin{equation}
	\cF_{rt} = q e^{\l \vf} r^{z-3}, \qq e^{\l \vf} = r^{\l \sqrt{4(z-1)}}, \qq q^2 = 2 \ell^2 (z+2) (z-1).\label{ex3mat} 
\end{equation}
The Lifshitz exponent is bounded from below by $z=1$, corresponding to the AdS black hole solution with no matter fields.

It turns out that one can obtain a stationary black hole solution by the following coordinate transformation \cite{Stachel:1981fg,Lemos:1994xp}
\begin{equation}
\dd{t} \rightarrow \Xi \dd{t} - a \dd{\theta}  \qq \qq \dd{\th} \rightarrow \frac{a}{\ell^2} \dd{t} - \Xi \dd{\th},\label{boost}
\end{equation}
where
\begin{equation}
	 \qq \Xi = \sqrt{1+\frac{a^2}{\ell^2}}, 
\end{equation}
and $a$ is a constant parameter. Note that this transformation is well-defined only locally but not globally due to the compactified nature of the $\th$-coordinate. If the first Betti number of a manifold is non-vanishing, which is one here, then there are no global diffeomorphisms that can map the original metric to the resulting metric. This means that the resulting manifolds must be parameterized globally by $a$. In \cite{Herrera-Aguilar:2021top}, the authors calculated the global charges explicitly and showed that one has a stationary black hole characterized by its mass and angular momentum, which is directly related to the parameter $a$. The resulting stationary line element is as follows
\begin{equation}
		\dif s^2=  - \frac{r^{2z}}{\ell^{2z} h(r)}  (\Xi \dif t-a \dif \theta)^2 + \frac{\ell^2 \dif r^2 }{r^2 h(r) }+\frac{r^2}{\ell^2} \dif x^2 + \frac{r^2}{\ell^4} \left(a \dif t- \ell^2 \Xi  \dif \theta\right)^2. 
	\label{rotmet}
\end{equation}
The transformation \eqref{boost} must also be applied to the matter fields\footnote{See \cite{Awad:2002cz} for an application to charged AdS black holes.} given in \eqref{ex3mat}, which has no effect on the scalar field $\vf$. However, we now have the following independent non-zero components of the Maxwell two-form $\cF_{\m\n}$
\begin{equation}
	\cF_{rt} = \Xi Q e^{-\lambda \vf} r^{z-2},\qq \cF_{r \theta}  =a \Xi Q e^{-\lambda \vf} r^{z-2}.
\end{equation}

In order to study the single copy properties, the metric should be first written in the KS coordinates. We will again take what we get from the asymptotic behaviour ($h \to 1$) as our background spacetime. Since it corresponds to the ``no black hole'' case, it is the appropriate ``flat spacetime limit'' of our solution, on which the Maxwell's equations should be defined. We take the line element of the background spacetime as
\begin{equation}
	\dif \bar{s}^2=  - \frac{r^{2z}}{\ell^{2z}}  (\Xi \dif t-a \dif \theta)^2 + \frac{\ell^2 \dif r^2 }{r^2}+\frac{r^2}{\ell^2} \dif x^2 + \frac{r^2}{\ell^4} \left(a \dif t- \ell^2 \Xi  \dif \theta\right)^2. 
	\label{rotmetback}
\end{equation}
Applying the transformation \eqref{boost} to the vector $k_\m$ of the static spacetime in \eqref{kandf}, we get
\begin{equation}
	k_\m \dd{x}^\m = \Xi \dd{t} + \frac{\dd{r}}{r^{z+1}}-a \dd{\theta},\label{rotkpre}
\end{equation}
which is null and geodesic with respect to the full metric \eqref{rotmet} and the background metric \eqref{rotmetback} as required. We will scale it such that $k_t =1$, which makes $V \cdot k = 1$, and use the following version
\begin{equation}
k_\m \dd{x}^\m = \dd{t} + \frac{\dd{r}}{ \Xi r^{z+1}}-\frac{a}{\Xi} \dd{\theta}.\label{rotk}
\end{equation}
As discussed in subsection \ref{subsec:KSDC}, this simplifies the resulting single and zeroth copy equations. The KS scalar $\f$ can be directly read as
\begin{equation}
	\f  = \Xi^2 \frac{r^{2z}}{\ell^{2z}}(1-h).
\end{equation}
As in the static case, the matter configuration is the same in the KS coordinates, which can be directly verified from the field equations. Because of the non-zero $\th$-component of the vector $k_\m$ in \eqref{rotk}, in addition to the electric field $E \equiv F_{rt}$, we also have a non-zero magnetic field $B \equiv F_{r \th}$. With the identification \eqref{Adef}, they read
\begin{equation}
	E = - \f^\pr, \qquad B = a \f^\pr = - a E.
\end{equation}
For our metric function in \eqref{ex3met}, one has
\begin{equation}
	\f = \Xi^2 \frac{\rp^{z+2} r^{z-2}}{\ell^{2z}}, \qq E = - \Xi^2 \frac{(z-2) \rp^{z+2}r^{z-3}}{\ell^{2z}}, \qq B = -a E,
\end{equation}
and the extra term $E^\mu$ in the Maxwell's equations \eqref{single} again vanishes since a zero vector transforms to itself under any coordinate transformation. As a result, we have a vacuum solution of the Maxwell's equation defined on the background spacetime with the line element in \eqref{rotmetback}.

For stationary type D spacetimes, the Weyl scalar $\Psi_2$ and the scalar $Z$ are in general complex (see \cite{Easson:2022zoh} where the Kerr-Newman black hole solution is studied). However, here we have a special case where they are real. They read
\begin{align}
	\Psi_2 &= \frac{1}{6} z \left(z-1\right)+\frac{ \left(z-4\right)  r_+^{z+2}}{6\, r^{z+2}}, \\
	Z &= \frac{\Xi  (z-2) r_+^{z+2}}{r^2}.\label{ex3Z}
\end{align}
This might be surprising especially for the scalar $Z$ since in general one expects that the magnetic field described by the single copy field is proportional to the imaginary part of $Z$. However, in this example, the field strength tensor in the flat frame, which is given by $F_{ab} = \bar{e}^\m_{\ a} \bar{e}^\n_{\ b} F_{\m\n}$ ($\bar{e}^\m_{\ a}$: vierbein of the background spacetime), has no magnetic component ($F_{12} = 0$ although $F_{r\th} \neq 0$). As a result, the field strength spinor, which is found by $f_{AB} = F_{ab} \s^{ab}_{AB}$ where $\s^{ab}_{AB}$ is the Infeld-van der Waerden symbol, takes the form in \eqref{fbasis} with a real $Z$, and as a result, the consistency condition in \eqref{CC} remains valid for this stationary black hole spacetime. 

Since the solution is valid as long as $z>1$, the vanishing of the relevant term in $\Psi_{2}$ and $Z$, corresponding to $z = 4, 2$,  should be studied separately. Note that these are the cases with the critical values $n_* = 2z-2$ and $n_* = 2z$ since $n=z+2$. Apart from the irrelevant $\Xi$ factor in the scalar $Z$ in \eqref{ex3Z}, the analysis is identical to that of the static black hole performed in \cite{Alkac:2023glx}. Therefore, we will not repeat it here. After regularizing $\Psi_{2}$ and $Z$ when required, the consistency condition is satisfied.

\section{Summary and outlook}\label{sec:sum}
In this work, we have examined the consistency between the KSDC and the SWDC, for a set of Lifshitz black hole solutions, each having a novel feature not studied before. Due to their anisotropic scaling properties, the Lifshitz black holes provide an ideal testing ground for the consistency of the two formulations of the CDC. The main reason for that is there exist many scenarios where quantities required to match the results obtained in the KSDC vanish and a regularization scheme should be applied to make them non-zero.

Our first example is a solution whose metric function is more complicated than the examples considered previously and have two terms with different origin that require regularization. The second example is a solution obtained by introducing an $R^2$-correction to the Einstein-Hilbert action. The last example is a special stationary black hole solution that is obtained from the static version by a coordinate transformation. For all these examples, the regularization scheme introduced in \cite{Alkac:2023glx} was found to be successful and the consistency of the KSDC and the WDC is achieved. 

While studying the KSDC of static Lifshitz black holes, we employed the coordinate transformation proposed in \cite{Ayon-Beato:2014wla}, which leads to the Lifshitz spacetime in \eqref{lif} as the background spacetime. This seems to be a quite natural choice since it is the ``zero-mass'' limit ($h \to 1$) of the black hole metric in \eqref{lifbh} and provides a straightforward generalization of the KSDC with a flat background. Therefore, we adopted the same strategy for the stationary black hole studied in subsection \ref{subsec:sta}. It is with these background metrics that we achieved the above-mentioned consistency. On the other hand, for AdS black holes, there appeared alternatives for the background spacetime recently \cite{Hassaine:2024mfs, Ayon-Beato:2025ahb}. Since the WDC calculations are performed in the Boyer-Lindquist coordinates, it is not obvious whether the consistency of the two formulations will survive when different backgrounds are used in the KS formulation\footnote{We thank the anonymous referee for raising a question regarding this point}. We expect that similar alternatives should also exist for Lifshitz black holes because they are generalizations of AdS black holes with an anisotropic scaling of the time coordinate. Therefore, it is of importance to study the alternative backgrounds and check the consistency also for these backgrounds.

Apart from this issue, the examples studied here provide further evidence for the validity of the regularization procedure. The essence of the regularization is that one gets the correct functional dependence in all the relevant quantities for a generic metric function of the form \eqref{h}, however sometimes one gets a zero contribution to the Weyl scalar due to the conformal flatness or to the $Z$ scalar due to the vanishing of the single copy electric field. Absorbing a zero factor in the coefficient of the relevant term, one recovers the consistency. While the form of the metric function we considered here is quite typical, for spacetimes with a line element different than \eqref{lifbh} of Lifshitz black holes, the Weyl scalar will take a different form. Also, the stationary solution that we considered here seems to have quite special properties since it is obtained by a coordinate transformation, which might not exist for a general stationary solution. It would be interesting to test the regularization scheme for a different type of black hole solution with a generic background metric, together with its stationary version if available. We leave this as future work.

\paragraph{Acknowledgments} This work was supported by T\"{U}B\.{I}TAK under Project Number 124F058. We thank G\"{o}k\c{c}en Deniz Al{\i}c{\i} for her help in various computations.

\bibliographystyle{utphys}
\bibliography{ref}

\end{document}